\documentclass[fleqn,twoside]{article}
\usepackage[headings]{espcrc2}

\readRCS
$Id: espcrc2.tex,v 1.2 2004/02/24 11:22:11 spepping Exp $
\usepackage{graphicx}
\usepackage[figuresright]{rotating}

\newcommand{\AmS}{{\protect\the\textfont2
  A\kern-.1667em\lower.5ex\hbox{M}\kern-.125emS}}

\title{Towards LHC phenomenology at the loop level:\\ 
       A new method for one-loop amplitudes}

\author{T. Binoth\address{School of Physics, University of Edinburgh, EH9 3JZ Edinburgh, UK}%
        \thanks{Work supported by Deutsche Forschungsgemeinschaft (DFG), grant number Bi1050/1-2.},
        M. Ciccolini\address{Paul Scherrer Institut, W\"urenlingen und Villigen, CH-5232 Villigen PSI, Switzerland}%
        and 
        G. Heinrich\address{Institute for Theoretical Physics, University of Z\"urich, CH-8057
	Z\"urich, Switzerland}%
	\thanks{Work supported by the Swiss National Science Foundation (SNF), contract number 200020-109162.}}
       
\runtitle{}

\runauthor{}

\begin{document}

\begin{abstract}
A precise understanding of LHC phenomenology requires the 
inclusion of one-loop corrections for multi-particle
final states. In this talk we describe a semi-numerical  
method to compute one-loop amplitudes with many
external particles and present first applications.
\vspace{1pc}
\end{abstract}

\maketitle

\section{Introduction}

\unitlength=1mm
\begin{picture}(0,0)
\put(120,110){
\begin{minipage}{5cm}
Edinburgh 2006/05\\
PSI-06-04\\
ZU-TH 05/06\\
January 2006\end{minipage}}
\end{picture}
In this decade high energy physics will explore the TeV scale
with hadron colliders like the Tevatron and the LHC.
The Tevatron experiments already have provided rich information
on electroweak, jet and heavy quark physics.
These data and their theoretical description are of major
importance for Higgs  and New Physics searches.
A detailed understanding of Standard Model processes is indispensable
to discriminate between signals and backgrounds
in this respect. This is especially true in hadronic collisions
at the TeV scale because of the complicated multi-particle final
states. It is well known that leading order
QCD predictions are generally plagued by large renormalization/factorization scale 
uncertainties. Only the inclusion of
higher order corrections may lead to more stable
predictions, as leading logarithmic contributions
typically cancel in that case. Moreover, it should be emphasized that not only 
% the stability under scale variations, respectively 
the overall normalization is an issue here, 
higher order corrections can also change the {\em shapes} of 
distributions. The knowledge of distribution shapes
is particularly important in those cases where backgrounds
measured in a low signal region have to be extrapolated 
to signal regions by using theoretical predictions.
A prominent example is the Higgs discovery mode
$H\to WW \to l\bar{\nu}\,\bar{l}'\nu'$.

The computation of one-loop corrections to multi-particle
final states is a complex task, as the combinatorial
growth of expressions together with complicated denominator
structures  hinders a straightforward  evaluation of the
amplitudes. 
Due to its phenomenological relevance, a lot of activity 
has been going on in this direction during the last years 
\cite{Denner:2005nn,Denner:2005fg,Anastasiou:2005cb,Ellis:2005zh,Binoth:2005ff,vanHameren:2005ed,Giele:2004iy,Nagy:2003qn,Binoth:2002xh,Duplancic:2003tv,Ferroglia:2002mz,Binoth:1999sp}.
 
In this talk, a method\,\cite{Binoth:2005ff}  for
the evaluation of one-loop diagrams  occurring in 
multi-particle computaions at NLO is presented. 
This approach allows for  
a good analytical control over the expressions, 
but also provides  stable numerical representations  
in phase space regions which typically pose numerical problems.   
Some applications are also presented.

\section{Reduction formalism}

Any given amplitude can be represented as a linear combination
of Feynman diagrams, which correspond to linear combinations
of tensor integrals. We define tensor integrals by 
\begin{eqnarray}
I^{n,\,\mu_1\ldots\mu_r}_N(a_1,\ldots,a_r) = &&\nonumber\\
\int d\bar{k}
\; \frac{q_{a_1}^{\mu_1}\,\dots  q_{a_r}^{\mu_r}}{
(q_1^2-m_1^2+i\delta)\dots (q_N^2-m_N^2+i\delta)}\;, &&
\end{eqnarray} 
where $q_i=k+r_i$, $r_{j}-r_{j-1}=p_j$, $\sum_{j=1}^N p_j=0$ and $d\bar{k}=d^nk/(i\pi)^{n/2}$.
One of the advantages of this representation is that the 
 combinations $q_i=k+r_i$ appear naturally (e.g. fermion propagators).
%Further it allows for a manifestly translation invariant formulation, means the
When introducing form factors, the Lorentz structure is carried by difference vectors 
$\Delta_{i\,j}^{\mu}=r_i^{\mu}-r_j^{\mu}$
and $g^{\mu\nu}$.
As an example, consider the rank 2 case 
\begin{eqnarray}
I_N^{n,\mu_1 \mu_2}(a_1,a_2;S)  
=  
\sum_{l_1,l_2 \in S}  \;
\Delta^{\mu_1}_{l_1 \, a_1} \;  \Delta^{\mu_2}_{l_2 \, a_2} \, 
A^{N,2}_{l_1 \, l_2}(S) &&\nonumber \\
+\; g^{\mu_1 \, \mu_2} \, B^{N,2}(S) \nonumber\;. &&
\end{eqnarray}
The kinematical information is encoded in the matrix 
${\cal S}_{ij} =  (r_i-r_j)^2-m_i^2-m_j^2$ which is the characteristic
object in any Feynman parameter integral. After momentum integration, 
the scalar $N$-point function is of the form
\begin{eqnarray*}
I^{n}_N(S) &=& 
(-1)^N\Gamma(N-\frac{n}{2}) \nonumber \\
 && \times \int \prod_{i=1}^N dz_i\,
\delta(1-\sum_{l=1}^N z_l)\,\left(R^2\right)^{\frac{n}{2}-N}\nonumber\\
&& R^2 =  
-\frac{1}{2} \sum\limits_{i,j=1}^N z_i\,{\cal S}_{ij} 
z_j\,-i\delta\;.
\end{eqnarray*}
Tensor integrals lead to additional factors $z_i$ in the numerator. 
As a shorthand notation we use the ordered set $S$ of propagator labels
instead of the  matrix ${\cal S}$ as our function arguments. 
A matrix corresponding to an integral with  propagator $1/P_j = 1/(q_j^2-m_j^2+i\delta)$
omitted (``pinched") corresponds to an index
set $S\setminus \{j\}$.
The reduction of scalar integrals can now be written as 
\begin{eqnarray}
I_N^n(S) &=& \sum_{i\in S} b_i(S) \, \int  d\bar{k} \;
\frac{ P_i}{\prod_{j\in S} P_j} \nonumber \\ && 
+ \int d\bar{k} \; 
\frac{ 1 - \sum_{i\in S}  b_i(S) \, P_i }{\prod_{j\in S}  P_j}\nonumber\\
& \stackrel{!}{=} & I_{div}(S) + I_{fin}(S)\;. \nonumber
\end{eqnarray}
One can show\,\cite{Binoth:2005ff} that
if  $\sum\limits_{i\in S} b_i(S)\,{\cal S}_{ij}=1$ is fulfilled, then
%\;(j\in S)
\begin{eqnarray*}
I_{fin}(S) & = & 
-B(S)\, (N-n-1)\, I_N^{n+2}(S)\;,\nonumber\\
%&&\nonumber\\
B(S)&=&\sum_{i\in S} b_i(S)\;, \nonumber\\
B(S)\,\det{\cal S}&=& (-1)^{N+1}\det G,\; G_{ij}=2\,r_i\cdot r_j\;,
\end{eqnarray*}
where the subscript ``{\em fin}'' indicates that the respective integral
is infrared finite. The algorithm allows to isolate
recursively  all infrared poles in terms of three-point integrals.

The reduction of tensor integrals is performed in
a similar way. We add and subtract a linear combination of
pinched terms and adjust the coefficients conveniently:
\begin{eqnarray}
&&I_N^{n,\mu_1\ldots\mu_r}(a_1,\ldots,a_r;S) =  \nonumber\\
& & \mbox{} - \sum_{j \in S}{\cal C}_{ja_1}^{\mu_{1}} \,  
\int d\bar k \;\frac{ P_j\;
q_{a_2}^{\mu_{2}}\ldots q_{a_r}^{\mu_{a_r}}}{
\prod_{i \in S} P_i}\nonumber\\
&&+\int d\bar k \, \frac{ \left[\,q_{a_1}^{\mu_{1}} + 
\sum_{j \in S} {\cal C}_{ja_{1}}^{\mu_{1}}\, P_j\right] \; 
 q_{a_2}^{\mu_{2}}\ldots q_{a_r}^{\mu_{r}}}{\prod_{i \in S} P_i} 
\nonumber\\
&& \stackrel{!}{=}  I_{div} + I_{fin} \nonumber
\end{eqnarray}
If  $\sum\limits_{i\in S} {\cal C}_{i\,a}^{\mu}(S)\,
{\cal S}_{ij}=\Delta_{j\,a}^\mu$ for all $j\in S$,  
then $I_{fin}$ is proportional to a sum of higher dimensional integrals 
$I_N^{n+2m}\,(m>0)$ and thus IR finite. In this way, the tensor integrals 
also can be reduced to a set of (potentially IR divergent) integrals 
with lower rank and less propagators and
an infrared finite part.
If ${\cal S}$ is invertible, i.e. for $N<7$ and non-exceptional kinematics,
one finds
\begin{eqnarray*}
%\sum\limits_{i\in S} {\cal C}_{i\,a}^{\mu}(S)\,{\cal S}_{ij}&=&\Delta_{j\,a}^\mu\\
%\Leftrightarrow\; 
{\cal C}_{i\,a}^{\mu}(S)&=&\sum_{j \in S} 
\left( {\cal S}^{-1} \right)_{ij} \Delta_{j a}^{\mu}\;,
\end{eqnarray*}
otherwise the pseudo-inverse $H_{ij}$ of the 
Gram matrix  $G$ can be used\,\cite{Binoth:2005ff,Binoth:1999sp} to obtain 
\begin{eqnarray*}
{\cal C}_{i\, b}^{\mu} &=& -\sum_{j\in S\setminus \{a\}} H_{ij}\,
\Delta_{j\, a}^{\mu} + W_i^{\mu}\,,\; i\in S\setminus \{a\}\\
{\cal C}_{a\, b}^{\mu} &=& 
-\sum_{j\in S\setminus \{a\}}{\cal C}_{j\, b}^{\mu}\;,
\end{eqnarray*}
where the vectors $W_i^{\mu}$ span the kernel of $G$. 

If $N> 5$, one finds that 
%${\cal V}_{b}^{\mu}\equiv 0$ and ${\cal T}^{\mu  \nu}_{a_1 \, a_2}\equiv 0$,
all higher dimensional integrals drop out as a consequence of 
rank\,$({\cal S}) = {\rm min}(N,6)$. 
Therefore one has for $N> 5$:
\begin{eqnarray*}
&&I_N^{n,\,\mu_1 \ldots \mu_{r}}(a_1, \ldots,a_r\,;S)=\nonumber\\&& -\sum_{j\in S} {\cal C}_{j \, a_{r}}^{\mu_{r}} \, 
I^{n,\,\mu_1  \ldots \mu_{r-1}}_{N-1}(a_1, \ldots,a_{r-1}\,;S 
\setminus\{j\}) \;.
\end{eqnarray*}
Note that $N$ and the tensor rank are reduced at the same time.
The case $N=5$ is more \,involved, but a reduction scheme has been worked out 
where no higher dimensional 
integrals $I_N^{n+2m} \,(m>0)$ are present for $N>4$
and  no inverse Gram determinants are introduced\,\cite{Binoth:2005ff}. 
As in the scalar case, an algebraic separation of IR poles, 
contained in 3-point integrals only, 
is achieved iteratively.

Application of the reduction formulas to integrals of rank $r\leq 2$
leads, for any $N$, to form factors in terms of simple scalar integrals.
For rank $r>2$, reduction to purely scalar integrals with $N=3,4$ is no longer possible
without introducing a tower of higher dimensional integrals. To avoid the latter, 
we use  integrals with Feynman parameters in the numerator as
reduction endpoints.
Explicit representations for all form factors for $r\leq N\leq 5$ 
can be found in\,\cite{Binoth:2005ff}.
The form factors are expressed in terms of the following basis integrals
\begin{eqnarray}
&&I^{n}_3(j_1, \ldots ,j_r) = 
-\Gamma \left(3-\frac{n}{2} \right)  \nonumber \\
&&\times\int_{0}^{1} \prod_{i=1}^{3} \, d z_i \, 
\, \frac{z_{j_1} \ldots z_{j_r}\delta(1-\sum_{l=1}^{3} z_l) }{ 
(-\frac{1}{2}\, z \cdot   {\cal S}
\cdot z)^{3-n/2}}\;,\label{i3r} \\
&&I^{n+2}_3(j_1) = 
-\Gamma \left(2-\frac{n}{2} \right) \nonumber \\&&
\times\int_{0}^{1} 
\prod_{i=1}^{3} \, d z_i \, 
\, \frac{z_{j_1}\delta(1-\sum_{l=1}^{3} z_l) }{ 
(-\frac{1}{2}\, z \cdot   {\cal S}
\cdot z)^{2-n/2}}\;, \nonumber\\
&& I^{n+2}_4(j_1, \ldots ,j_r) = 
\Gamma \left(3-\frac{n}{2} \right) \nonumber \\&& \times
\int_{0}^{1} \prod_{i=1}^{4} \, d z_i \,  
\, \frac{z_{j_1} \ldots z_{j_r}\delta(1-\sum_{l=1}^{4} z_l)}{ 
(-\frac{1}{2}\, z \cdot   {\cal S}
\cdot z)^{3-n/2}}\;, \label{i4r}\\
&&I^{n+4}_4(j_1) =
\Gamma \left(2-\frac{n}{2} \right) \nonumber \\ 
&&\times\int_{0}^{1} \prod_{i=1}^{4} \, d z_i \,  
\, \frac{z_{j_1}\delta(1-\sum_{l=1}^{4} z_l)}{ 
(-\frac{1}{2}\, z \cdot   {\cal S}
\cdot z)^{2-n/2}}\;,\nonumber
\end{eqnarray}
where $r_{\rm max}=3$ in eqs.\,(\ref{i3r}) and (\ref{i4r}),  
and of course purely scalar integrals $I^{n}_3,I^{n+2}_3,I^{n+2}_4,I^{n+4}_4$
are also basis integrals.
We provide two alternatives for the evaluation of the basis integrals.
On one hand, an algebraic reduction to scalar integrals can be performed.
This leads to representations with
inverse Gram determinants. As long as the  Gram determinant -- or more precisely
the related quantity $B\sim  \det G/\det {\cal S}$ --
is sufficiently large, such a representation is numerically safe.
On the other hand, if the Gram determinants become smaller
than  a certain cut parameter $\Lambda$, a direct numerical evaluation
of the building blocks of the reduction is performed.

\section{Numerical evaluation of  basis integrals}

\begin{figure}[h]
%\vspace{9pt}
%\framebox[55mm]{\rule[-21mm]{0mm}{43mm}}
%\includegraphics[width=12.0cm]{TRGfig7.ps}
\unitlength=1mm
\begin{picture}(70,70)
\put(-30,-45){\includegraphics[width=11.5cm]{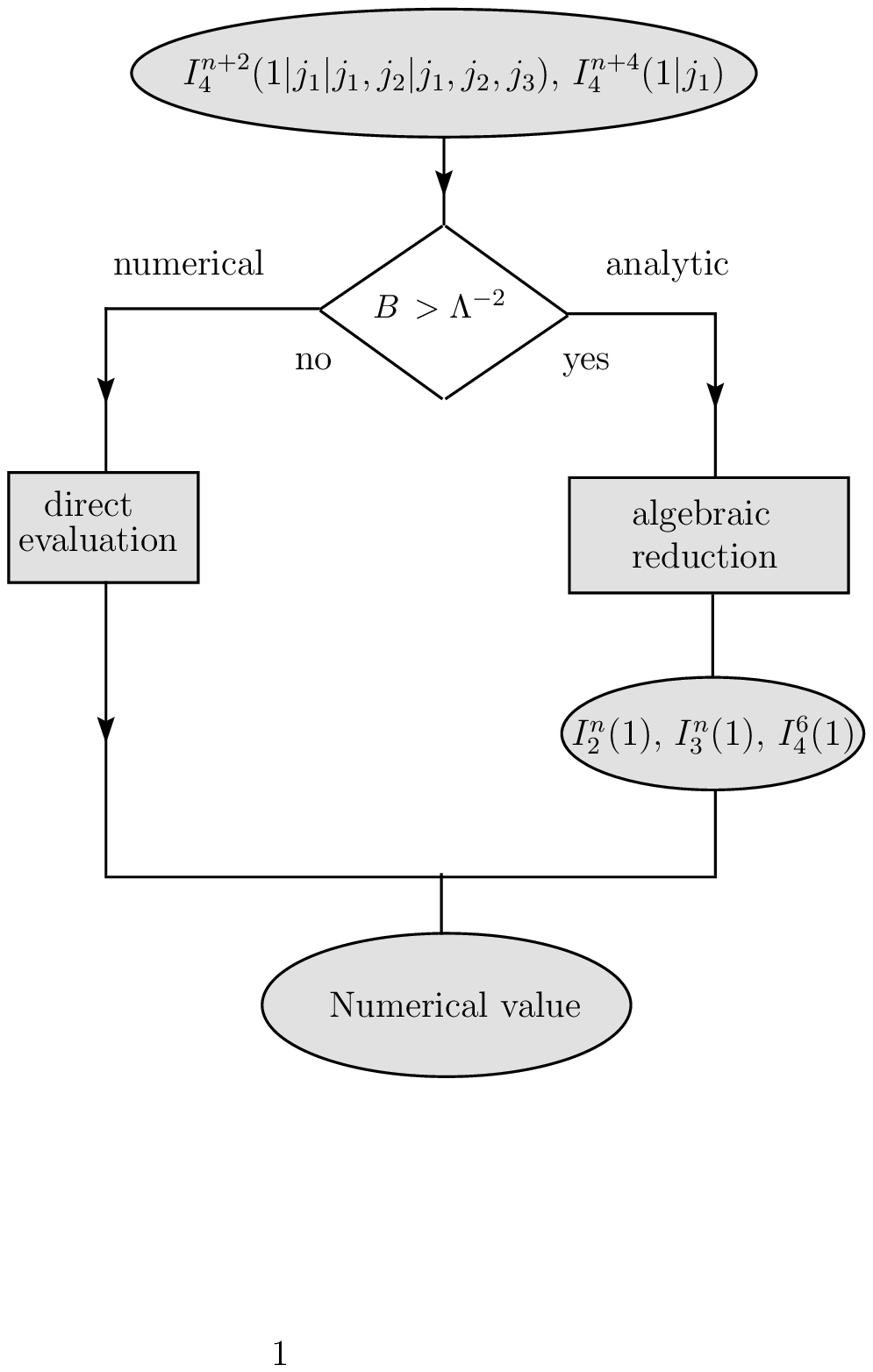}} 
\end{picture}
\caption{Schematical evaluation of basic box functions. $\Lambda$ 
is a user-defined parameter serving as a switch between analytic/numerical
representations.}
\label{fig:schema}
\end{figure}

We show now how a numerically stable evaluation of the integrals
\begin{eqnarray}
&&I_N^{d}(j_1,\dots ,j_r) =  (-1)^N\Gamma(N-\frac{d}{2}) \nonumber\\
&&\int_{0}^{\infty} d^Nx\,\delta(1-\sum\limits_{l=1}^N x_{l})
\frac{x_{j_1}\dots x_{j_r}}{(x\cdot {\cal S}\cdot x/2 +i\delta)^{N-d/2}} \nonumber
\end{eqnarray}
for the required cases $N\leq 4$, $r\leq 3$, 
$d=4-2\epsilon,6, 8-2\epsilon$
can be achieved. 
Let us focus on the basis integrals with $d=4-2\epsilon$, $N=3$ and $d=6$, $N=4$.
The other cases contain UV divergencies which are isolated in terms of 
external $\Gamma(N-d/2)$ factors. After expansion
of the integrand in $\epsilon$, the pole part is a trivial integral
and the finite part is a logarithmic integral which can be treated
with the same contour deformation methods as  discussed below.
IR divergencies are  only present  for $N=3$ and $d=4-2\epsilon$. 
For the IR divergent integrals 
explicit analytical formulas can be found in \cite{Binoth:2005ff}.
The numerical evaluation of the remaining 
IR finite integrals is problematic due to 
kinematical singularities, which occur 
if the quadratic form $x\cdot {\cal S}\cdot x$ changes sign. 
We propose the following solution:
%To eliminate the $\delta(1-\sum x_l)$ 
We first make a sector decomposition
\begin{eqnarray}
1 = \sum\limits_{l=1}^N 
\theta( x_l > x_1,\dots,x_{l-1},x_{l+1},\dots,x_N )\;.\nonumber
\end{eqnarray}
The integral is thus split into $N$ terms:
\begin{eqnarray}
&&I_N^d(j_1,\dots ,j_r) = (-1)^N \Gamma(N-d/2) \nonumber\\
&& \sum\limits_{l=1}^N  J_l(N,d,j_1,\dots ,j_r)\;.\nonumber
\end{eqnarray}
In sector $l$ one applies the variable transformation $x_j=t_j x_l$ for $(j<l)$, $x_j=t_{j-1}x_l$ 
for  $(j>l)$ and integrates out $x_l$ with the $\delta$ distribution. 
Defining $\vec T=(t_1,\dots,t_{l-1},1,t_{l},\dots,t_{N-1})$ gives
\begin{eqnarray}
&&J_l(N,d,j_1,\dots ,j_r) = \int\limits_0^1 d^{N-1}t \;
\Bigl( \sum\limits_{j=1}^N T_j \Bigr)^{N-d-r}\nonumber \\
&&\times\frac{T_{j_1}\dots T_{j_r}}{ \Bigl( T\cdot {\cal S}\cdot T/2 - 
i\delta \Bigr)^{N-d/2}}\nonumber\;.
\end{eqnarray}
$Q(t)=T\cdot {\cal S}\cdot T/2$ leads to singular behaviour if
\begin{eqnarray}
Q(t) = \frac{1}{2} \sum\limits_{j,k=1}^{N-1} A_{jk} t_j t_k + 
\sum\limits_{j=1}^{N-1} B_{j} t_j + C = 0\nonumber
\end{eqnarray}
$A,B,C$ are defined by ${\cal S}$.
For the sector  integrals $J_l$ the following 
contour deformation,  
parametrized by $\alpha$, $\beta>0$, $\lambda\geq 0$,
leads to smooth integrands \cite{Binoth:2005ff,Kurihara:2005ja}
\begin{eqnarray}
\vec{x}( \vec t) &=& \vec t - i\;  \vec \tau(\vec t)\nonumber\\
\tau_k &=& 
\lambda \, t_k^\alpha\, (1-t_k)^\beta 
\sum\limits_{j=1}^{N-1} ( A_{jk} t_j +B_k  ) \nonumber
\end{eqnarray}
While 
$\lambda \nabla\cdot Q$ controls the size of the deformation, 
the parameters $\alpha,\beta$ control the
smoothness of the deformation at the integration boundaries. 
There are situations where the contour deformation
is not possible. This happens if $\nabla\cdot Q=0$ and $Q=0$. 
This exceptional kinematic situation 
occurs in the presence of normal or anomalous thresholds
and cannot be avoided. The way to deal with
it is to split the integrations at
$t_j = - \sum\limits_{l=1}^{N-1}  A^{-1}_{jl} B_l$
if necessary. Typically the proposed contour deformation method
works well, although the CPU time for evaluating 
a basis integral is much larger than evaluating the corresponding
analytical representations. The conclusion for the practitioner
is to use the fast and accurate algebraic formulas for the ``bulk"
of the phase space and switch to the slow but reliable 
numerical evaluation near critical phase space regions.
In the case of 4-point functions this looks schematically as shown in Fig.~\ref{fig:schema}.
%\unitlength=1mm
%\begin{picture}(70,60)
%\put(-35,-60){\includegraphics[width=12.0cm]{TRGfig7.ps}} 
%\end{picture}

\vspace*{-3mm}

\section{Applications}
\begin{figure}[h]
%\vspace{9pt}
%\framebox[55mm]{\rule[-21mm]{0mm}{43mm}}
%\includegraphics[width=12.0cm]{TRGfig7.ps}
\unitlength=1mm
\begin{picture}(80,55)
\put(0,-8){\includegraphics[width=7.5cm]{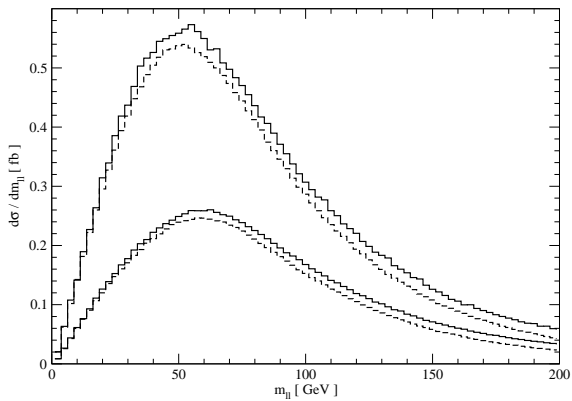}} 
\end{picture}
\caption{Invariant mass distribution of the charged lepton pair for
$gg\to W^*W^*\to l\bar{\nu}\,\bar{l}'\nu'$. The two sets of curves are
the LHC prediction with (lower) and without (upper) standard cuts. 
The effect of the third generation massive quark loop leads to a slight
enhancement (full) compared to the case with two massless generations (dashed). 
For parameter choices see \cite{Binoth:2005ua}.}
\label{ggww}
\end{figure}

Our reduction algorithm has been applied so far to several
processes \cite{Binoth:2001vm,Binoth:2002qh,Binoth:2003xk,Binoth:2005ua}. 
Here we focus on two of them. First, 
off-shell vector boson pair production via gluon fusion at the LHC, 
which is a  
background for the $gg\to H \to W^*W^*$ channel \cite{Binoth:2005ua}.
Although only box graphs occur, the large number of
invariants $(s,t,p_3^2,p_4^2,m_b,m_t)$ makes the process
sufficiently involved to provide a good testing
ground for our method. 
The helicity amplitudes $\Gamma^{++}$, $\Gamma^{+-}$ were computed in a modular way.
The amplitude was decomposed into gauge invariant terms and was
reduced completely to $d=6$ box, $d=4$ triangle and bubble integrals.
%Stable numerical representation for $m_q=0$ and $m_q\neq 0$ could be
%obtained.
As an illustration we show the invariant mass distribution 
of the charged lepton pair for 
$gg\to W^*W^* \to l\bar{\nu}\, \bar{l}'\nu'$ in Fig.~\ref{ggww}.
Numerically instable regions are confined to very small
phase space regions which do not contribute to the result inside the 
errors. 

%\vspace*{-3mm}

Another application of our reduction procedure
was the first direct computation of the $gg \to \gamma\gamma g$ one-loop amplitude.
Although the amplitude could be indirectly deduced from the $gg\to ggg$ loop
result \cite{Bern:1993mq}, it was
shown that very compact results could be obtained for
all six independent helicity amplitudes \cite{Binoth:2003xk}.
%$\Gamma^{+++++}$, $\Gamma^{++++-}$,
%$\Gamma^{+++--}$, $\Gamma^{+-+++}$, $\Gamma^{+-++-}$, $\Gamma^{--+++}$ 
This result is of relevance for the Higgs boson search channel
$PP\to H+\mbox{jet}\to \gamma \gamma \;\mbox{jet}$ at the LHC.

\section{Summary}
For a precise understanding of LHC phenomenology, NLO precision
for  multi-particle amplitudes is mandatory. We have
presented a  
new semi-numerical approach for 1-loop multi-leg processes
which is valid for an arbitrary number of massless as well as massive 
internal/external particles.
It allows to isolate IR divergences 
transparently and in an automated way from the amplitude. 
%It was designed to respect manifestly translation invariance.
Our reduction algorithm uses a set of building blocks
which can be evaluated either analytically, if 
the inherent Gram determinants are sufficiently large, 
or numerically, avoiding inverse Gram determinants and the associated 
instabilities completely.  
To achieve the latter, a
multi-dimensional contour deformation method for 
1-loop parameter integrals was developed.
The full potential of the contour deformation method for 
one-- and multi-loop Feynman diagrams 
is being further investigated.

We have shown that our analytical reduction method works well for
processes of phenomenological interest like photon pair plus jet production 
or off-shell vector boson pair production at the LHC.
Applications to LHC processes with a
larger number of external particles are presently being studied.
% are in progress. 

\end{document}